# Dimensionality controlled Mott transition and correlation effects in single- and bi-layer perovskite iridates

Q. Wang,[1] Y. Cao,[1] J. A. Waugh,[1] S. R. Park,[1] T. F. Qi,[2] O. B. Korneta,[2] G. Cao,[2] and D. S. Dessau[1]

[1]Department of Physics, University of Colorado, Boulder, CO 80309-0390, USA

[2]Center for Advanced Materials, Department of Physics and Astronomy, University of Kentucky, Lexington, KY 40506, USA

We studied $Sr_2IrO_4$ and $Sr_3Ir_2O_7$ using angle-resolved photoemission spectroscopy (ARPES), making direct experimental determinations of intra- and inter-cell coupling parameters as well as Mott correlations and gap sizes. The results are generally consistent with LDA+U+Spin-orbit coupling (SOC) calculations, though the calculations missed the momentum positions of the dominant electronic states and neglected the importance of inter-cell coupling on the size of the Mott gap. The calculations also ignore the correlation-induced spectral peak widths, which are critical for making a connection to activation energies determined from transport experiments. The data indicate a dimensionality-controlled Mott transition in these 5d transition-metal oxides (TMOs).

Compared to the extensively studied 3d TMOs, such as high-$T_c$ cuprate superconductors (1) in which the strong electron correlation plays a dominant role in determining the electronic structures, the 5d TMOs have several fundamental differences: the 5d electrons are more extended in real space which leads to a large band width and a reduced Coulomb correlation, and the very large atomic number leads to a large SOC effect. The delicate interplay between electron correlations, SOC, inter-site hopping, and crystal field splitting leads to a strongly competing ground state for the 5d TMOs, including the iridates. So far, a great amount of exotic physics behaviors have been theoretically proposed to exist in the iridates, such as high-$T_c$ superconductivity (2), quantum spin Hall and correlated topological insulator effects (3-6), and a Weyl Fermion state (7), though these are all currently lacking experimental verifications. Recently, the insulating behavior of single- and bi-layer perovskite strontium iridates $Sr_2IrO_4$ and $Sr_3Ir_2O_7$ has been explained by the cooperative interplay between correlation effects and strong SOC of the iridium 5d electrons (8-13). The optical conductivity, ARPES, X-ray absorption spectroscopy, and resonant inelastic X-ray scattering (8,9,13) all appear consistent with these materials being classified as $J_{eff}=1/2$ Mott insulators, though a recent theoretical proposal (14) claimed that they are actually Slater insulators.

In this letter, we report a systematic ARPES study on single- and bi-layer perovskite iridates $Sr_2IrO_4$ and $Sr_3Ir_2O_7$. The band dispersions of both materials were mapped and compared to the available calculations. While the overall electronic structures of both materials appear roughly consistent with the LDA+U+SOC calculations based on the $J_{eff}=1/2$ Mott ground state picture, important differences remain. Specifically, we found the lowest energy (closest to $E_F$) states locate near the zone corner (X point), while they are theoretically predicted to be at the $\Gamma$ point. This may be due to an underestimate of the SOC strength, or a strongly momentum-dependent electronic self-energy. An additional aspect missing from the calculations is the three-dimensional (3D) inter-cell coupling, which is significant (~ 100 meV) in $Sr_3Ir_2O_7$ and almost absent in $Sr_2IrO_4$. This inter-cell coupling appears to drive the reduction in the Mott gap, placing $Sr_3Ir_2O_7$ on the precipice of a Mott transition. Finally, the finite spectral peak widths are completely ignored in the calculations and are argued to be highly relevant for making a connection to transport experiments as well as for predicting when the Mott transition may occur.

High-quality single crystals of $Sr_2IrO_4$ and $Sr_3Ir_2O_7$ were synthesized using a self-flux technique (15,16). The crystals were cleaved in situ and measured in an ultra-high vacuum better than $3 \times 10^{11}$ torr. The ARPES experiments were performed at Beamline PGM-A (071) at the Synchrotron Radiation Center (SRC), Madison, and Beamline 7.0.1 at the Advanced Light Source (ALS), Berkeley. The angular resolution of the experiments was approximately $0.1^o$ and the energy resolution was 20 ~ 35 meV (depending upon photon energy).

Fig. 1(a) shows the in-plane crystal structure of $Sr_2IrO_4$ and $Sr_3Ir_2O_7$. An important structural feature of these compounds is that they crystallize in a reduced tetragonal structure due to a rotation of the $IrO_6$-octahedra about the c-axis by ~ $11^o$, resulting in a larger in-plane unit cell by $\sqrt{2} \times \sqrt{2}$ as shown as the red dashed box in the figure (17,18). Figs. 1(b1)-1(b4) show the intensity maps of $Sr_2IrO_4$ at different binding energies from 0.1 eV to 0.4 eV. The white box in panel (b1) shows the first two-dimensional (2D) Brillouin zone (BZ) boundary with high symmetry points labeled. Consistent with its insulating behavior, there is no spectral weight at the Fermi level. Figs. 1(c1)-1(c5) show the spectra taken along the high symmetry directions over several BZs as indicated by the yellow cuts in panel (b1). Figs. 1(d1)-1(d5) are the second-derivative images along the energy direction of spectra (c1)-(c5), respectively, which enhances the contrast of the raw spectra and makes it easier to track the electronic dispersion. The solid red lines are guides to the eye for the dispersions that can be resolved in the spectra, while the dashed red lines are guides to the eye for the dispersions that cannot be resolved in the specific spectra due to the matrix element effect (19,20) but do show up at the spectra taken along the same high symmetry direction but at different BZs. This dispersion data should be viewed as the centroids of spectral weight, as it is sometimes not possible for us to individually distinguish multiple bands that are close together. Figs. 2(a1)-2(a5) show the intensity maps of $Sr_3Ir_2O_7$ from the Fermi level to 0.4 eV. Compared to the single-layer material there is slightly more spectral weight at the Fermi level, which is principally due to the increased "leakage" of spectral weight up to the Fermi level due to the smaller gap of $Sr_3Ir_2O_7$. This leakage can be due to intrinsic correlation effects as well as the finite energy resolution of the experiment, and will be discussed again in conjunction with fig. 5. Figs. 2(b1)-2(b5) show the spectra taken along the high symmetry directions over several two-dimensional BZs as indicated by the yellow cuts in panel (a1). Similar to $Sr_2IrO_4$, there is no band crossing the Fermi level but the dispersions are much closer to the Fermi level. So the spectral weight shown in fig. 2(a1) rather indicates a smaller energy gap in $Sr_3Ir_2O_7$ than in $Sr_2IrO_4$. Figs. 2(c1)-2(c5) are the second-derivative images along the energy direction of spectra (b1)-(b5), respectively. Again, the solid and dashed lines in panels (c1)-(c5) are guides to the eye for the experimentally observed dispersions.

Fig. 3 shows a compilation of the dispersion data for $Sr_2IrO_4$ and $Sr_3Ir_2O_7$, as well as a comparison to the LDA+U+SOC calculations adopted from S. J. Moon *et al* (9). Figs. 3(b) and 3(d) are the experimentally extracted in-plane dispersions for both materials, which again should be viewed as the centroids of spectral weight. First of all, for both materials, there is no band crossing the Fermi level, which is consistent with their insulating behavior. Compared to the single-layer compound, the uppermost band of bilayer $Sr_3Ir_2O_7$ is much closer to the Fermi level. This is much clearer in figs. 3(e) and 3(f), which show stacks of EDCs along the Γ-X-Γ line for both materials. Furthermore, there are clearly more bands observed for $Sr_3Ir_2O_7$ in the same energy range, which is naturally explained as the bilayer splitting due to the intra-cell coupling. In particular, the bilayer splitting is observed to be minimum at the X point and maximum at the Γ point, with magnitude about 0.25 eV.

Fig. 3(a) and 3(c) are the theoretical calculations, in which a U value of 2.0 eV and a SOC constant of 0.4 eV were used to optimize the calculation for matching to optical conductivity spectra. In the calculations, as the result of strong SOC, the Ir 5d $t_{2g}$ band splits into the effective $J_{eff}=1/2$ (doublet) and $J_{eff}=3/2$ (quartet) bands. The near-$E_F$ half-filled $J_{eff}=1/2$ band further splits into the effective upper and lower Hubbard bands due to its very small effective band width in spite of the relatively small on-site Coulomb repulsion. Here we note that the overall band calculations match the experimentally determined dispersion reasonably well for both materials, with no shifting or scaling of the data. Consistent with the lack of scaling, the intra-cell bilayer splitting observed in the calculation of $Sr_3Ir_2O_7$ (~ 0.25 eV) is also fully consistent with the experiment. This lack of a scaling is surprising for correlated electron materials, which are usually found to have reduced bandwidths (or enhanced masses) relative to the LDA calculations. The lack of a scaling found here may be a result of the relatively small U for these materials. The overall agreement between the band calculation and the experimental dispersion provides additional evidence for this "$J_{eff}=1/2$ Mott state" picture, and encourages us to follow the theoretical calculation and color code our experimental result with red dashed lines representing the $J_{eff}=1/2$ bands (21).

Despite the significant agreement between the theory and experiment, important differences remain. In the calculations, the lowest energy occupied states are at Γ (at both Γ and X for $Sr_2IrO_4$), as highlighted by the red ovals. As seen from the plots of fig. 3, the bilayer splitting at Γ is the reason why $Sr_3Ir_2O_7$ is predicted to have a larger bandwidth and smaller gap than $Sr_2IrO_4$. In contrast, for the experiment the states at Γ are farther away from the Fermi level, and the states at X will be the most dominant for the low energy properties. Figs. 4(a)-(c) are LDA, LDA+spin-orbit coupling (SOC), and LDA+SOC+U band calculations of $Sr_2IrO_4$, adopted from B. J. Kim *et al* (8). Fig. 4(d) is the experimental dispersions of $Sr_2IrO_4$ obtained by our ARPES measurement. Based on the calculations, the introduction of SOC (~ 0.4 eV) will introduce the splitting of the $t_{2g}$ bands into $J_{eff}=1/2$ and $J_{eff}=3/2$ bands. From the calculated dispersions, this splitting is significant at both the Γ and M points, while it is minimal at the X point. At the Γ point, the blue and red bands in panel (a) split into the purple and pink bands in panel (b), with the purple band moving to lower energy. At the M point, the highly degenerate group of bands at -0.5 eV splits, with two of the bands moving towards $E_F$, as indicated by the up-arrow. The on-site correlation effect (~ 2 eV) further splits these two bands, with the splitting almost uniform in momentum space as shown in panel (c). In an overall picture, from the calculations, the SOC in the system makes the lowest energy electronic state near Γ move away from $E_F$ while doing the opposite at the M point. By increasing the SOC strength parameter used in the calculations, we may expect that the calculated lowest energy electronic states near Γ and M will keep moving towards the opposite direction and at a certain level the electronic state near Γ will have a higher binding energies than the state at M (and X). This is exactly what we obtained from the experiment as shown in panel (d). Due to the overall agreement between the band calculations and the experimental dispersions, in panel (d) we color-coded the experimental dispersions in a similar way as shown in the calculations. It shows that the lowest energy electronic state near Γ has larger binding energy than the state near M, as indicated by the arrows at those two high symmetry points. This is an indirect evidence of the underestimation of the SOC in the calculations. Alternatively, the disagreement in the energy positions of the bands between experiment and theory may be explainable as a strongly momentum-dependent self-energy.

Based on the discussion above, the intra-cell bilayer splitting is not seen to affect the bandwidth in the actual material, and so it is not expected to directly affect the magnitude of the Mott gap. In the calculation, the individual $J_{eff}=1/2$ bands in $Sr_2IrO_4$ are positioned at the average energy of the bilayer split $J_{eff}=1/2$ bands of $Sr_3Ir_2O_7$. Therefore, both materials show the same energy gap at the X point (where the bilayer splitting goes to zero). Another mechanism is therefore needed to explain the difference in the gap size between the two materials. We will show that this is most likely the inter-cell coupling, which we experimentally find to be significantly larger for the bilayer samples. Additionally, we note that in the calculation, the same U is used for both materials. Hence this discrepancy may indicate a reduced correlation effect in bi-layer materials which may be due to the feedback effect, where the extra metallicity in $Sr_3Ir_2O_7$ due to the intra- and inter-cell coupling partially screens the onsite U and results in a reduced correlation effect.

Fig. 5 shows details of the states at the X point (the zone corner of the 2D BZ), including the inter-cell coupling effects. Inter-cell coupling gives rise to a coherent dispersion perpendicular to the planes, which can be accessed by varying the incident photon energy. Such data are shown in figs. 5(a) and 5(b), which are stacks of EDCs at the zone corner with photon energy varying from 80 eV to 140 eV. Here we find that throughout the large photon energy range, $Sr_2IrO_4$ always shows a larger energy gap than $Sr_3Ir_2O_7$. Furthermore, for $Sr_2IrO_4$ the EDCs at the zone corner taken with different photon energies all have a very similar lineshape and peak position. In contrast, for $Sr_3Ir_2O_7$ the EDC lineshapes change a lot as a function of excitation energy and the peak position shows very strong photon energy dependence. Figs. 5(c) and 5(d) summarize the low energy peak positions of $Sr_2IrO_4$ and $Sr_3Ir_2O_7$ as a function of photon excitation energy. Utilizing the free-electron final-state approximation, we can convert the photon energy to a corresponding $k_z$ value by $k_z = \sqrt{\frac{2m}{\hbar^2}(E_i + h\upsilon - \phi - V_0) - k_{||}^2}$ for both $Sr_2IrO_4$ and $Sr_3Ir_2O_7$, where $h\upsilon$ is the photon excitation energy, $\phi$ is the sample work function, and $V_0$ is an experimentally determined inner potential (23). By fitting the periodicity of the spectra, the inner potentials of -27.6 eV for $Sr_2IrO_4$ and -25.4 eV for $Sr_3Ir_2O_7$ are obtained. Here we note that for $Sr_2IrO_4$ the experimental result shows the periodicity as $4\pi/c_1$, where $c_1$ = 25.8 Å as the lattice constant in the c direction. This is fully consistent with its tetragonal lattice structure. On the other hand, for $Sr_3Ir_2O_7$ the experimental dispersion shows the periodicity close to $8\pi/c_2$, where $c_2$ = 20.9 Å is its c-axis lattice constant. This extra factor of two in the periodicity is not understood. Nevertheless, by defining the energy gap as the peak-to-$E_F$ distance, we can parameterize the gap size for both materials as the function of $k_z$, obtaining $\Delta_{n=1} = 0.287 -$

$0.005\cos(k_z)$ and $\Delta_{n=2} = 0.136 + 0.039\cos(k_z)$, both in eV. The black dashed lines in figs. 5(c) and 5(d) represent the fitting results. Here we plot them as a function of excitation energy instead of $k_z$ value since the $k_z$ is essentially a function of photon excitation energy. The high symmetry points along z direction are labeled as $\Gamma$ and Z for $Sr_2IrO_4$, and $\Gamma'$ and Z' for $Sr_3Ir_2O_7$. These data allow us to find the absolute minimum of the peak in the 3D BZ, which is 97 meV from $E_F$ for $Sr_3Ir_2O_7$ and 282 meV from $E_F$ for $Sr_2IrO_4$. We see that the increased dimensionality in $Sr_3Ir_2O_7$ significantly affects the Mott gap, because the inter-cell $k_z$ dispersion acts directly on the lowest energy states at the X point.

The sharpest leading edge for $Sr_3Ir_2O_7$ is of order 90 meV, and the sharpest edge for $Sr_2IrO_4$ is more than 180 meV. In both cases this is much larger than the experimental energy resolution, which is at most 35 meV. Therefore, these edge widths and pole energies should have very minimal shifts due to resolution effects, confirmed by our simulation also (not shown). On the other hand, the very small but still finite spectral weight observed at $E_F$ in $Sr_3Ir_2O_7$ is within the resolution window of the experiment, i.e. it is possibly fully a result of the finite experimental resolution. Because the pole energy is not affected by the finite energy resolution we can make firm connections between this energy scale and other experimental probes, in particular transport. Within the standard theory of insulators, the electrical resistivity relates to a so-called activation energy $E_A$ through an exponential relation as $\rho \sim \rho_0 \cdot \exp(E_A/k_BT)$ where the activation energy $E_A$ has a value of half of the band gap ($E_A = E_g/2$), and the band gap $E_g$ is the energy difference between the occupied and unoccupied quasiparticle poles. Typically we expect the Fermi energy to be near the center of the band gap – in that approximation the

ARPES pole energies would predict resistive activation energies of 282 meV for $Sr_2IrO_4$ and 97 meV for $Sr_3Ir_2O_7$. These values are much larger than the measured activation energies from transport, which are roughly 105 meV for $Sr_2IrO_4$ (22) and 20 meV for $Sr_3Ir_2O_7$ (16). Even if we make the most extreme approximation that the unoccupied pole is exactly at the chemical potential (which would then give a clear signal in the occupied ARPES spectral weight) we predict activation energies of 141 meV for $Sr_2IrO_4$ and 48 meV for $Sr_3Ir_2O_7$, which are between 40% and 150% larger than the actual value from transport. This disagreement is highlighted in figs. 5(e) and 5(f), where the experimental activation energies are plotted on top of the ARPES data. As just discussed, these differences are so large that they can't be due to the possible uncertainty of the Fermi energy location within the band gap, but rather must be due to a fundamental breakdown of the simple picture that relates the pole energies to the activation energies. Rather, the data of figs. 5(d) and 5(e) indicate that the activation energy is related to the onset of the spectral weight of the pole, rather than the pole energy itself. Though such a direct comparison to activation energies has not to our knowledge been made before, such behavior has been observed in other correlated electron systems, for example the manganites (24) or cuprates (25) where the peak energy is observed to be far from $E_F$, argued in both cases to be the result of correlation effects (in particular polaronic effects). Such a distinction of course only makes sense for a correlated electron insulator with a finite peak width, as a standard insulator should have vanishingly small peak width for the low energy peak in the low temperature limit (in the same way that the low temperature quasiparticle width of a Fermi liquid metal goes to zero at $E_F$). The fact that the less correlated $Sr_3Ir_2O_7$ has a smaller peak width than the more correlated $Sr_2IrO_4$ is also consistent with this behavior. Going beyond $Sr_3Ir_2O_7$, we envision that a further increase in the dimensionality could bring about a further reduction in the gap energies such that the edges of the spectral peaks could overlap with $E_F$ while the peak centroids remain away from $E_F$. Such a metal, where the spectral peaks potentially never reach the Fermi energy, would have analogies to the famous pseudogap states in the manganites and cuprates, further cementing the similarities between different classes of Mott insulators on the verge of metallicity.

In summary, by using ARPES, we studied the electronic structure of single- and bi-layer perovskite iridates $Sr_2IrO_4$ and $Sr_3Ir_2O_7$. The overall electronic structures of both materials are partially consistent with the LDA+U+SOC calculations based on the $J_{eff}=1/2$ Mott ground state picture, though the calculations also miss some critical physics. The different dimensionality between these two materials, in particular, the strong intercell coupling in $Sr_3Ir_2O_7$ makes it have less Mottness with smaller energy gap, sharper peaks, and larger $k_z$ dispersion comparing to $Sr_2IrO_4$.

This work was supported by the National Science Foundation under grant DMR-1007014 to the University of Colorado and grants DMR-0856234 and EPS-0814194 to the University of Kentucky. This work is also based in part upon research conducted at the Advanced Light Source, which is funded by the US Department of Energy, and at the Synchrotron Radiation Center which is primarily funded by the University of Wisconsin-Madison with supplemental support from the University of Wisconsin-Milwaukee.

*Note added:* after finishing our work, we become aware of recent ARPES studies on $Sr_3Ir_2O_7$ which reveal the in-plane electronic structures of $Sr_3Ir_2O_7$ that is basically consistent with our observation (26).

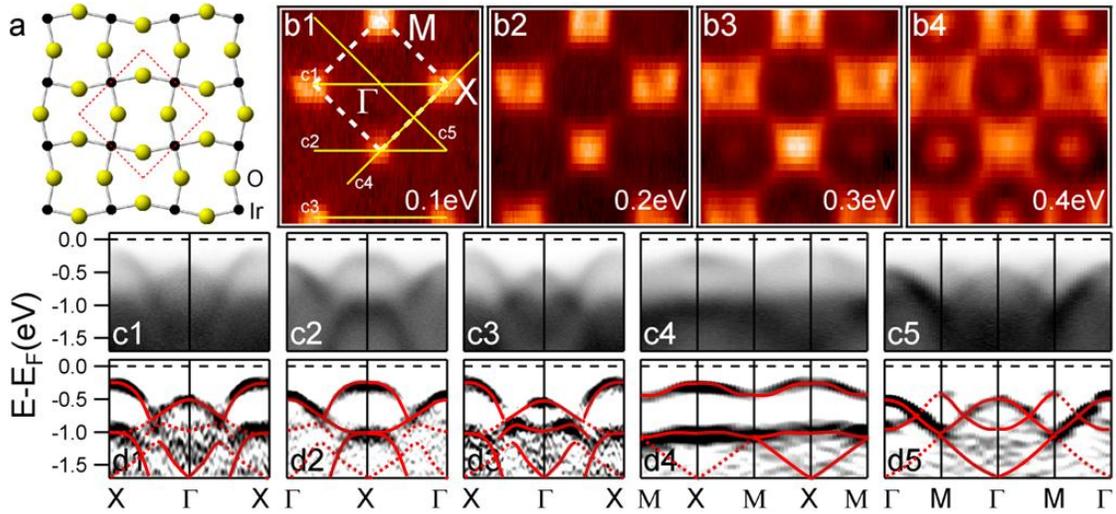

FIG. 1: (a) In-plane crystal structure of $Sr_2IrO_4$. The white dashed box represents the in-plane unit cell. (b1-b4) Intensity maps at different binding energies from 0.1 eV to 0.4 eV of $Sr_2IrO_4$. (c1-c5) Spectra taken along high symmetry cuts c1 to c5 as indicated by the yellow lines in panel (b1). (d1-d5) Second-derivative images along energy direction of spectra c1-c5, respectively. The red solid and dashed lines are guides for the eyes of the experimental observed dispersion. All data were taken with 80 eV photons at 25 K.

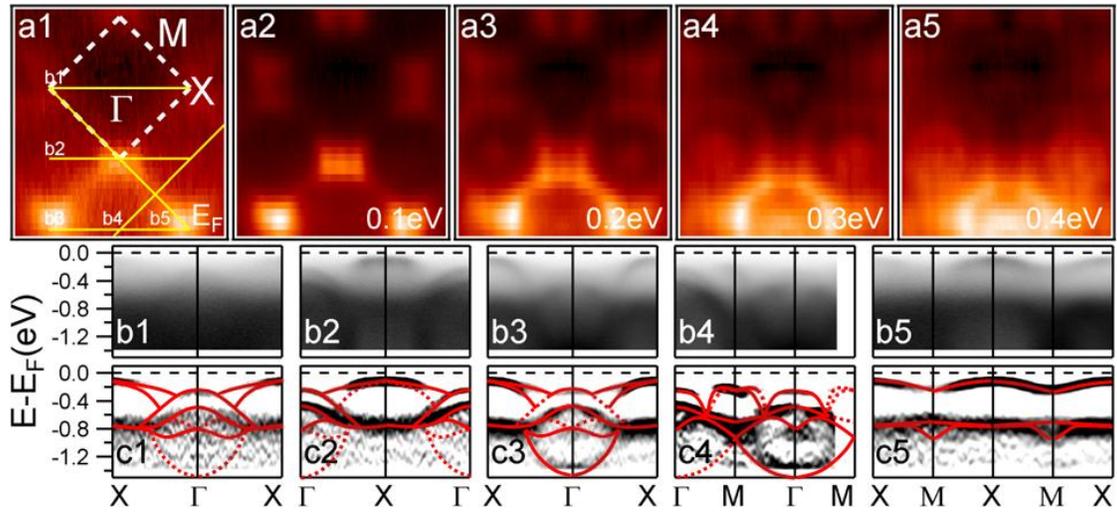

FIG. 2: (a1-a5) Intensity maps at different binding energies from Fermi level to 0.4 eV of $Sr_3Ir_2O_7$. (b1-b5) Spectra taken along high symmetry cuts b1 to b5 as indicated by the yellow lines in panel (a1). (c1-c5) Second-derivative images along energy direction of spectra b1-b5, respectively. The red solid and dashed lines are guides for the eyes of the experimental observed dispersion. All data were taken with 80 eV photons at 25 K.

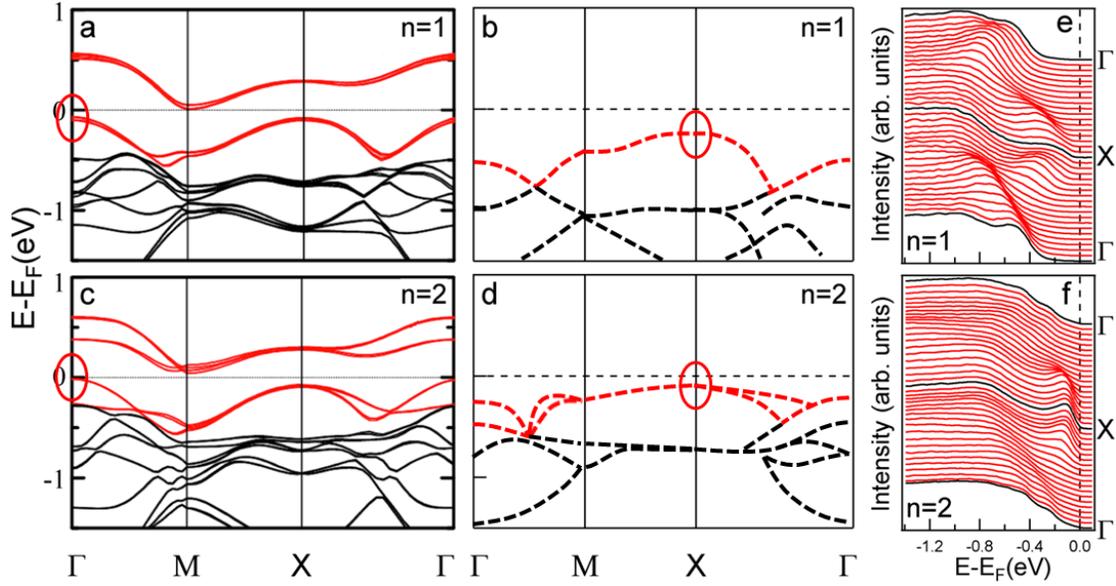

FIG. 3: (a,c) LDA+U+SOC band calculations of $Sr_2IrO_4$ (top) and $Sr_3Ir_2O_7$ (bottom), adopted from S. J. Moon *et al* (9). (b,d) Experimental dispersions (centroids of spectral weight) of $Sr_2IrO_4$ (top) and $Sr_3Ir_2O_7$ (bottom). The red and black lines represent the $J_{eff}=1/2$ and $J_{eff}=3/2$ bands, respectively. In the calculations, the dominant low energy occupied states are at the $\Gamma$ point (red ovals). In the experiment, the dominant low energy states are at the X point (red ovals) instead. (e) and (f) Stacks of EDCs along the $\Gamma$-X-$\Gamma$ directions for $Sr_2IrO_4$ (top) and $Sr_3Ir_2O_7$ (bottom).

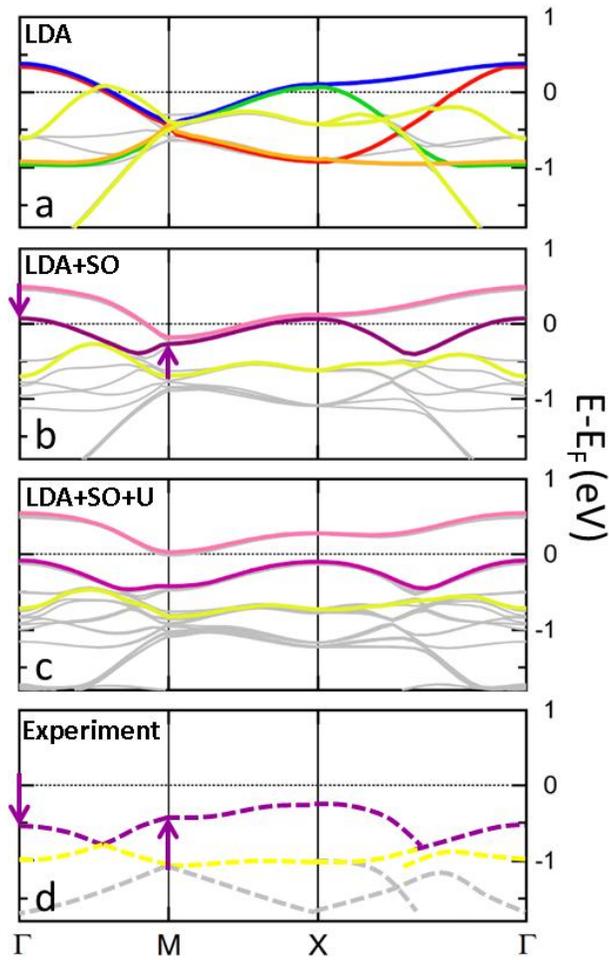

FIG. 4: Comparison between the experimental dispersions and the theoretical calculations of $Sr_2IrO_4$. Theoretical band dispersions of $Sr_2IrO_4$ in (a) LDA, (b) LDA+SOC (~ 0.4 eV), (c) LDA+SOC+U (~ 2 eV), adopted from B. J. Kim *et al* (8). (d) Proposed experimental dispersions of $Sr_2IrO_4$ for comparison. The vertical arrows in panel (b) show the clear impact of SOC on the band structure, which raises the low energy state at M and lowers it at $\Gamma$. As shown by the longer arrows in panel (d), the experimental data follows a trend that may be explainable with a still larger value of SOC.

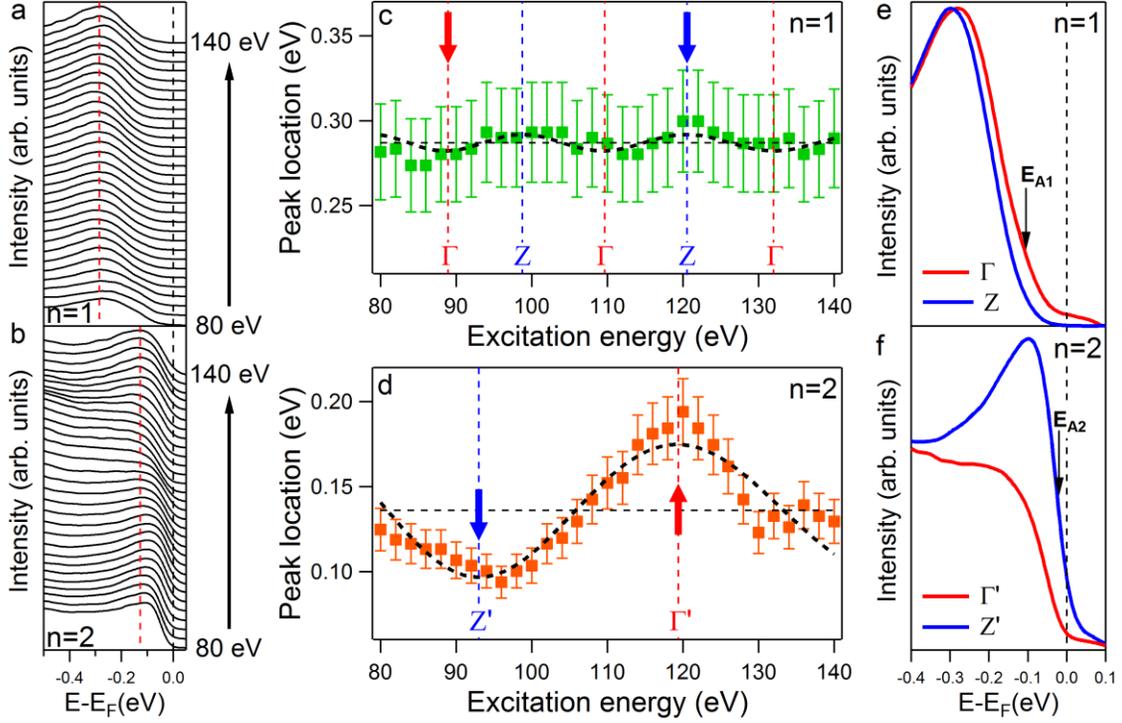

FIG. 5: (a) and (b) Photon energy dependence of the EDCs at the X point for $Sr_2IrO_4$ and $Sr_3Ir_2O_7$ (80 eV to 140 eV with 2 eV/step). The red dashed lines are guide to the eyes for viewing the variation of the EDC peak locations. (c) and (d) Extracted peak energy as a function of excitation energy for $Sr_2IrO_4$ and $Sr_3Ir_2O_7$, respectively. The black dashed curves in (c) and (d) are fitted curves obtained by fitting the peak energy into function $\Delta = \Delta_0 + \eta \cdot \cos(k_z)$, where $k_z = \sqrt{\frac{2m}{\hbar^2}(E_i + h\upsilon - \phi - V_0) - k_{\parallel}^2}$. The $\Gamma$ and Z labeled in (c), and the $\Gamma$' and Z' labeled in (d) indicate the high symmetry points in $k_z$ direction of $Sr_2IrO_4$ and $Sr_3Ir_2O_7$, respectively. (e) and (f) EDCs at $\Gamma$ and Z for $Sr_2IrO_4$, at $\Gamma$' and Z' for $Sr_3Ir_2O_7$, indicated by red and blue arrows in panel (c) and (d), respectively. The black arrows indicate the measured activation energies from transport for $Sr_2IrO_4$ ($E_{A1}$=105 meV) and $Sr_3Ir_2O_7$ ($E_{A2}$=20 meV).

**References**

1. J. G. Bednorz, and K. A. Muller, *Z. Phys. B* **64**, 189-193 (1986).

2. F. Wang, and T. Senthil, *Phys. Rev. Lett.* **106**, 136402 (2011).

3. A. Shitade *et al.*, *Phys. Rev. Lett.* **102**, 256403 (2009).

4. B. J. Yang, and Y. B. Kim, *Phys. Rev. B* **82**, 085111 (2010).

5. H. C. Jiang, Z. C. Gu, X. L. Qi, and S. Trebst, *Phys. Rev. B* **83**, 245104 (2011).

6. C. H. Kim, H. S. Kim, H. Jeong, H. Jin, and J. Yu, *Phys. Rev. Lett.* **108**, 106401 (2012).

7. X. G. Wan, A. M. Turner, A. Vishwanath, and S. Y. Savrasov, *Phys. Rev. B* **83**, 205101 (2011).

8. B. J. Kim *et al.*, *Phys. Rev. Lett.* **101**, 076402 (2008).

9. S. J. Moon *et al.*, *Phys. Rev. Lett.* **101**, 226402 (2008).

10. B. J. Kim *et al.*, *Science* **323**, 1329 (2009).

11. H. Jin, H. Jeong, T. Ozaki, and J. Yu, *Phys. Rev. B* **80**, 075112 (2009).

12. H. Watanabe, T. Shirakawa, and S. Yunoki, *Phys. Rev. Lett.* **105**, 216410 (2010).

13. K. Ishii *et al.*, *Phys. Rev. B* **83**, 115121 (2011).

14. R. Arita, J. Kunes, A. V. Kozhevnikov, A. G. Eguiluz, and M. Imada, *Phys. Rev. Lett.* **108**, 086403 (2012).

15. G. Cao, J. Bolivar, S. McCall, J. E. Crow, and R. P. Guertin, *Phys. Rev. B* **57**, R11039 (1998).

16. G. Cao *et al.*, *Phys. Rev. B* **66**, 214412 (2002).

17. M. K. Crawford *et al.*, *Phys. Rev. B* **49**, 9198 (1994).

18. Q. Huang *et al.*, *J. Solid State Chem.* **112**, 355 (1994).

19. Z.-X. Shen, and D. S. Dessau, *Phys. Rep.* **253**, 1-162 (1995).

20. A. Damascelli, Z. Hussain, and Z.-X. Shen, *Rev. Mod. Phys.* **75**, 473-541 (2003).

21. There is still some theoretical uncertainty over the nature of the states near the Γ point. In particular, the calculations done by Watanabe *et al*. indicated that the $J_{eff}=1/2$ and $J_{eff}=3/2$ states cross near the Γ point so that the $J_{eff}=3/2$ states are closer to $E_F$ (12). Further efforts are required to clarify the exact nature of these states.

22. M. Ge *et al*., *Phys. Rev. B* **84**, 100402(R) (2011).

23. S. Hüfner, *Photoelectron Spectroscopy* (Springer, 1995).

24. D. S. Dessau *et al*., *Phys. Rev. Lett.* **81**, 192 (1998).

25. K. M. Shen *et al*., *Science* **307**, 901 (2005).

26. B. M. Wojek et al., *J. Phys.: Condens. Matter* **24**, 415602 (2012).